# Adoption of the open access business model in scientific journal publishing – A cross-disciplinary study

Bo-Christer Björk and Timo Korkeamäki

Bo-Christer Björk is Professor of Information Systems Science at the Hanken School of Economics in Helsinki, Finland. Timo Korkeamäki is the Dean of the Business School at the Aalto University in Espoo, Finland. Emails: bo-christer.bjork@hanken.fi, timo.korkeamaki@aalto.fi.

## Abstract

Scientific journal publishers have over the past twenty-five years rapidly converted to predominantly electronic dissemination, but the reader-pays business model continues to dominate the market. Open Access (OA) publishing, where the articles are freely readable on the net, has slowly increased its market share to near 20%, but has failed to fulfill the visions of rapid proliferation predicted by many early proponents. The growth of OA has also been very uneven across fields of science. We report market shares of open access in eighteen Scopus-indexed disciplines ranging from 27% (agriculture) to 7% (business). The differences become far more pronounced for journals published in the four countries, which dominate commercial scholarly publishing (US, UK, Germany and the Netherlands). We present contrasting developments within six academic disciplines. Availability of funding to pay publication charges, pressure from research funding agencies, and the diversity of discipline-specific research communication cultures arise as potential explanations for the observed differences.

## Introduction

The Internet has been the catalyst for radical new business models across numerous fields and industries. Also, the publication process of scholarly journals has experienced changes in the past twenty-five years that are more significant than those during the prior three centuries combined, even though articles themselves look much the same as before. The primary technical change has been migration from print to electronic delivery, which has made changes in business models possible. Traditionally, the primary means of collecting revenue for journal publication has been subscription income from readers and from the libraries that serve them. The conversion to electronic publishing allows a potential reversal of the income model so that either authors pay for dissemination or the costs are met by subsidies from universities or scholarly societies. The end result of such a reversal is that peer-reviewed articles are made freely available for anyone with Internet connection, with no need for payments or registration. This phenomenon has been named Open Access[1]. OA is consistent with the fundamental ethos of science[2]. It makes scientific results easily accessible, it is cost efficient, and extremely convenient for potential readers. Through increased readership, OA increases the impact of science, regarding both the academic impact on the development of science itself, and the societal impact on the activities of other stakeholders in society[3, 4].

Open access comes in a number of variations. An article can be openly available at the journal website ("gold OA"), or the author can make an earlier manuscript version available in an open repository ("green OA")[5]. At the journal level, there are further distinctions. A subscription journal can allow authors to pay to make individual articles OA. Such journals are called hybrid OA journals. Other journals may make their full content available after a delay of typically one year, but require subscriptions for immediate access[6]. In this article, we focus only on journals which make all articles available OA immediately, which entails totally refraining from subscription income. Furthermore, full OA journals can either be newly created journals, which are typically electronic only, or established subscription-based journals, which have converted to Open Access[7]. In both cases, the publisher either charges authors so-called APCs (article processing charge), or finds another means to subsidize costs, including voluntary labor and open source software.

Had the OA uptake been dependent only on newly founded journals, the development in market share would have been slower than the 1% per year experienced during the past two decades. The conversion of existing journals to OA, by making the electronic version free and often continuing to sell print versions to subscribers, has contributed an almost equal share of the OA journals. Converted journals have a significant competitive advantage over newly founded journals since they have established prestige, and editorial networks. The most common case is that such journals continue to send print copies to subscribers but make the OA version free, where the OA publishing is subsidized by either the subscription income, or for instance via society membership fees. In some countries such as Canada, Norway, and Finland, specific funding sources exist in support of the publishing of scholarly journals and in some cases also the conversion to OA. Such funding is more common for society and university publishers. The case of subscription journals converting to electronic only distribution with APC funding is less common.

## Evolution of Open Access publishing

In the 1990's, OA journals were typically new journals, founded by independent academics on web sites they created themselves, and such journals where not to be found in recognized journal indexes. From 2003 onwards, the Directory of Open Access Journal (DOAJ) has provided a reasonable means of tracking the growth of OA journals. One way to measure the progression of the OA business model is to observe the proportion of OA journals or articles within the set of journals indexed in either the Web of Science or Scopus. Studies conducted at different points in time are not fully consistent but provide an indication of the development of the market share among qualitatively better scholarly OA journals.

The earliest study of this nature was McVeigh's 2004 study of the share of OA journals and articles in WoS[8]. She found 2.6 % of journals and 3.0 % of articles to be OA. Solomon et al studied the OA-journals indexed in Scopus retrospectively between 1999 and 2010 and found a journal share of 10.1 % for 2010[9]. More recently, Jubb et al report the share of OA-journals in Scopus in 2012 to be 12.4 % and the article share of 10.3 %[10]. The corresponding figures for 2016 are 15.2 % and 18.9 %, respectively. Our own observations in this study for 2017 indicate 18.4 % for journals and 18.9 % for articles. Combining the results from these studies for the share of OA journals of all journals (Figure 1) indicates a steady linear growth of the market share, with the shares having risen by approximately 1 percentage point per year.

Figure 1. Longitudinal development of the share of OA journal articles in the SCOPUS index.

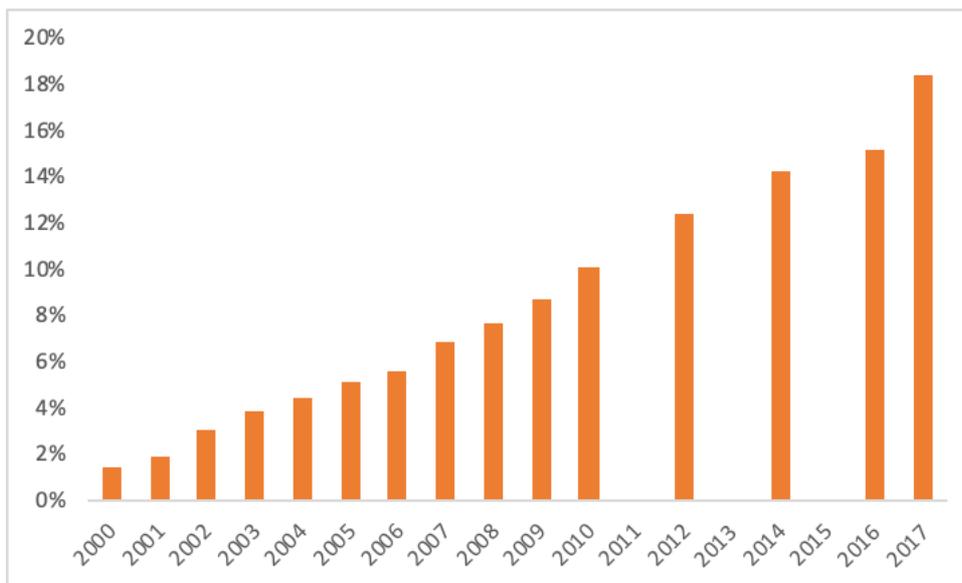

Excluded from these figures are the vast majority of journals from so-called predatory OA publishers. Such journals are mostly not included in Scopus, Web of Science (WoS) or the Directory or Open Access Journals (DOAJ).  Most academics receive spam email soliciting contributions from such journals on a daily basis. Several studies have demonstrated the lack of proper peer review by such publishers, in particular the experiment of journalist John Bohannon[11], which received a lot of media attention worldwide. They are nevertheless using the same business model as more quality-driven OA publishers. Predatory journals appear to cater to a segment of authors for whom just adding a published title to their CVs and rapid publication is sufficient. There are currently more than 11,000 journals listed in Cabell's blacklist of predatory journals and their annual article output was estimated to be 420,000 in 2014[12].

The growth in prevalence of OA publishing has not been uniform across different geographic/language market segments. Uptake has, for instance, been strongest in countries where English is not the dominant publishing language. In the Spanish-speaking world, the conversion of existing society and university journals has been strongly supported by free or extremely low-cost OA portals, such as Scielo or Redalyc[13]. Also, non-commercial publishers have been more active to convert than their commercial counterparts. Non-profit OA portals, often using the Open Source OJS software have played a major role. Many newly founded commercial OA publishers have profited from locating their operations in lower cost countries such as Egypt and China, enabling them to keep the APCs at competitive levels.

A number of OA prevalence studies have also included estimates of the uptake in different disciplines. Some have looked at gold OA in particular[14],[15], others at Gold OA as part of the OA availability in total [16], [17], [18]. All studies have showed big differences in uptake between disciplines. OA has gained a foothold more rapidly in the hard STM (Science, Technology & Medicine) sciences, in particular in biomedicine, and it has been much more difficult to get acceptance for this new business model in the social sciences and the arts and humanities.

***Our primary aim in this study is to provide a nuanced and updated picture of variation in OA adoption across disciplines. We use recent data and provide descriptive statistics of factors that potentially affect the uptake level in different academic disciplines. Such discipline-specific factors include the geographic variation in the origin of journal publishers, the OA share among top ranked journals, the share of commercially published journals, the shares of converted vs. born OA journals, and whether OA journals charged authors for publishing.***

***In addition to statistical comparisons, we select a few disciplines with contrasting OA developments as cases and discuss possible explanations for the developments based on available literature.***

## Methods and data

We conduct both quantitative and qualitative analysis of differences in OA prevalence between different academic disciplines. Our quantitative analysis is an extension of methods used in earlier studies [19], [20], [21]. This is followed by a qualitative consideration of potential explanations for the inter-disciplinary variation, using existing scholarly literature, reports and discussions in the media.

Studies of the total number of Open Access journals and their relative share of all scholarly journals tend to rely on different journal directories in order to identify all relevant peer reviewed journals, whether subscription or OA. Common data sources include the Web of Science (WoS), Scopus and Ulrich's, with WoS being the most restrictive regarding inclusion of journals. The WoS and Scopus also index individual articles, while Ulrich's only lists journals. The Directory of Open Access Journals (DOAJ) is a moderated index of Open Access Journals, the majority of which are indexed by neither WoS nor Scopus. DOAJ recently tightened its quality control in order to exclude predatory journals. The Directory of Open Access Scholarly Resources (ROAD) is managed by the ISSN organization, and it lists a slightly greater number of OA journals and periodicals than DOAJ[22].

Scopus and Ulrich's allow searches for journals using various criteria, including country of the publisher, citedness of the journals (i.e., impact factor), etc. As we are interested in the variation of OA prevalence between disciplines, classification of journals in different subject fields is of particular importance. The classification schemes used in the major indexes (WoS, Scopus, DOAJ) are however unfortunately not directly compatible with each other, which means that combining journal information from the three sources is difficult at the discipline level. Due to its broader coverage of journals (compared to WoS) and inclusion of article level data, Scopus is used as the basis in our study. Scopus has a three-tier classification structure, which can be easily searched using the free Scimago web site (https://www.scimagojr.com/). The second tier level is particularly useful for this study, with 27 categories. Scopus classifies some journals in multiple subject categories creating some overlap, but that overlap is generally limited and thus it has minimal impact on our analysis. However, the two biggest so-called OA megajournals which both publish tens of thousands of articles per year pose a problem. *PLOS ONE* (21,139 articles in 2017) is classified in three different categories "Medicine", "Biochemistry, Genetics and Molecular Biology" and "Biochemistry, Genetics and Molecular Biology". *Scientific Reports* (24,827 articles) is only classified under "Multidisciplinary", along with subscription journals such as *Science*, and *Nature*. Since *PLOS ONE* has a similar broad scope as the other three journals, we reclassify it as Multidisciplinary and remove it from the three more specialized subjects above. This gives us a more accurate picture of the share of OA articles across different fields.

Scopus has recently introduced tagging journals as Open Access journals based on information obtained from DOAJ and also ROAD. OA journals can be easily filtered on the Scimago site. Table 1 provides information regarding the total number of journals and OA journals in each index as of December 2018.

Table 1. Number of journals overall and OA journals in the leading indexing services.

| Index | No of journals | No of OA journals | OA journal share |
| --- | --- | --- | --- |
| WoS | 16,257 | 2,786 | 17.1 % |
| Scopus | 24,385 | 4,485 | 18.4 % |
| Ullrich's | 82,559 | 16,224 | 19.7 % |
| DOAJ |  | 12,135 | 100 % |
| ROAD |  | 16,224 | 100 % |
| Cabell's |  | 10,352 | 100 % |

The number of indexed journals varies greatly among the subjects on the second tier of the Scopus classification. The biggest field, Medicine, contains over 7,175 journals whereas Dentistry has only 196 journals. Also, Social Science is a very broad field with 5,716 journals but it excludes the business disciplines and psychology, which are closely related to the social sciences.

Since it is difficult to draw reliable estimates of OA shares from small categories with just a few hundred journals overall and hence only a handful of OA journals, we exclude the ten smallest second-tier categories in Scopus (with between 196 and 740 journals each). The journals in these categories stand for approximately 12% of all journals and some of them are also classified in other categories. Despite the comparatively low number of journals in the category "Multidisciplinary," we however retain this category due to the substantial number of articles in that category, mainly from the leading megajournals.

Another interesting aspect is the country of publication of the journals. Four countries totally dominate the global scholarly journal scene, in particular for English-language STM (Science, Technology & Medicine) subjects [23]. A number of leading scholarly society publishers and university presses, as well as several big commercial publishers, come from the United States and the United Kingdom. The Netherlands and Germany are home to the two leading commercial publishers, Elsevier and Springer Nature. In total, 63% of the 24,385 journals indexed in Scopus are published in one of these four countries. We will refer to UK, US, Germany and the Netherlands as the "Big four" through the remainder of the paper. There are several indications that OA publishing is less common among the leading subscription publishers located in these four countries. In contrast, prior studies indicate significantly higher OA shares for other countries and for publications in languages other than English[24], [25]. This detail motivates us to split the data into two world regions (the Big four countries vs. the rest of the world). This same split has earlier been used in a slightly different context by Solomon et al [26].

In order to account for the scientific quality of journals, we use a simplified method. Based on Scopus citation data, the Scimago service calculates a metric (Scimago Journal Rank, SJR) for the impact of the average article in a journal. The SJR accounts for both the number of citations received by a journal and the importance or prestige of the journals where such citations come from. Using the SJR score ranking within each discipline, we calculate the share of OA journals among the top 10 % of journals.

We also utilize a detailed dataset about 4,548 OA journals indexed in Scopus, provided by Mikael Laakso from ongoing research. These data contain an additional classification of the OA journals into those that have been OA from the start ("born-OA") versus subscription journals that have converted to OA. The dataset further contains a classification of the type of publisher (commercial, society or university) for each journal as well as data about whether a journal charges for publishing or not.

Longitudinal studies of OA developments using data from Indexes like Web of Science or Scopus are challenging to carry out. While obtaining article numbers for individual journals for past years is straight-forward, issues such as the point in time when a converted OA journal in fact turned OA are difficult to determine. For these reasons, we conduct only a cross-sectional study, based on data at the end of 2017.

## Quantitative results

Table 2 provides percentages of OA journals and articles that are published in OA journals, by discipline. The OA article shares are slightly lower than the OA journal shares for most disciplines, with the exceptions of Physics and Astronomy and Multidisciplinary journals.

Table 2. Uptake levels for OA journals and OA articles in them for 18 Scopus disciplines

|  | No of journals in Scopus | No of OA journals of all journals % | No of OA Articles of all articles % |
|---|---|---|---|
| All subject fields | 24,358 | 18.4 | 18.8 |
|  |  |  |  |
| Agriculture & Biological Sciences | 2,062 | 27.2 | 20.9 |
| Multidisciplinary | 114 | 24.3 | 75.3 |
| Biochemistry, Genetics and Molecular Biology | 2,002 | 23.4 | 23.0 |
| Medicine | 7,175 | 22.1 | 22.0 |
| Environmental Science | 1,344 | 20.2 | 14.5 |
| Earth and Planetary Sciences | 1,126 | 20.2 | 15.0 |
| Computer Science | 1,491 | 16.2 | 13.9 |
| Social Sciences | 5,716 | 15.6 | 14.6 |
| Mathematics | 1,382 | 14.4 | 13.7 |
| Physics and Astronomy | 1,039 | 14.0 | 15.6 |
| Materials Science | 1,138 | 13.9 | 9.0 |
| Chemistry | 802 | 13.8 | 13.9 |
| Engineering | 2,692 | 13.4 | 10.4 |
| Arts & Humanities | 3,570 | 12.9 | 11.2 |
| Economics, Econometrics and Finance | 941 | 11.8 | 8.0 |
| Psychology | 1,114 | 11.5 | 10.2 |
| Business | 1,230 | 7.6 | 6.0 |

Part of the inter-disciplinary differences in OA uptake may be explained by the geographical variation in publishing in the disciplines. While only 9% of journals for publishers located in the region composed of the four countries US, UK, Germany and Netherlands are OA, the average share is 34 % for all other countries combined, with Latin America's 80% at the top in a more detailed breakdown. For instance, in Agriculture & Biological Sciences, which has the highest OA journal share worldwide (27%), the share of journals published outside the Big four is also highest with 45 %. At the other end of the spectrum, in business, with the lowest total OA share (8 %), the share of journals published outside the Big four is only 18%.

For this reason, a more meaningful comparison between disciplines is to look at journals published in the Big four countries and in other countries, separately. For instance, for biochemistry the figures are 17 and 37%, and for business 2 and 31%. The differences between fields thus becomes much more pronounced when the figures for journals from the Big four countries and all other countries are separated as in figure 2 below.

Figure 2. The shares of OA journals in different scientific disciplines for journals published in different regions. The disciplines are ordered according to the percentage in the four leading publishing countries.

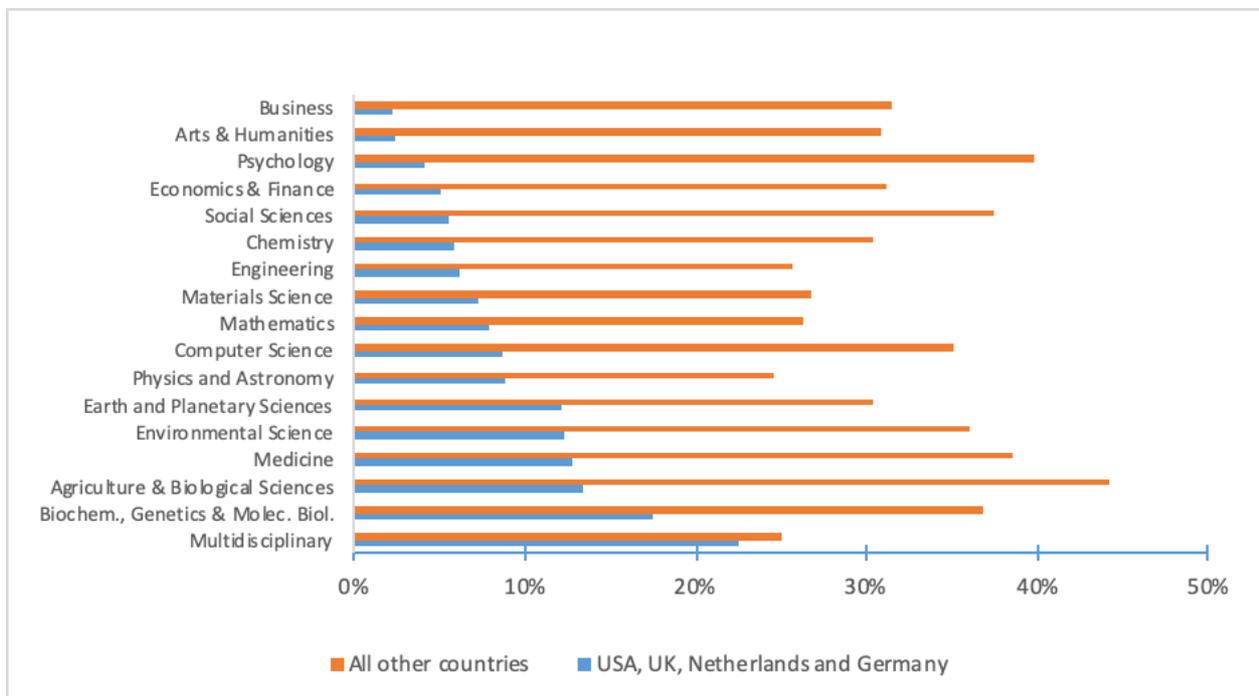

The figures for "all other countries" show less marked differences between disciplines in the share of OA, with a range between 25% and 44%. Also, they rank differently. For instance, psychology has a higher journal uptake (39,8 %) than biomedicine (36,8 %) and Physics and astronomy has the lowest (24,5) of all disciplines. It seems evident that discipline has much less importance in the uptake level of OA in other countries compared to the Big four.

It is also interesting to observe the differences between disciplines according to a number of OA related factors. We provide such a comparison in Table 3.

Table 3. Share of OA journals, share of top ranked OA journals according to the SJR citation metric, share of journals published by commercial publishers, share of born-OA journals, share that charge authors and the median APC for such journals.

| Subject Field | OA share of all journals | Top 10 % SJR share | Commer-cial share | Born OA share | Share APC journals | Median APC USD |
|---|---|---|---|---|---|---|
| All Subject Fields | 18.4 | 13.0 | 41.2 | 44.1 | 37.3 | 1,021 |
| | | | | | | |
| Agriculture & Biological Sciences | 27.2 | 20.4 | 28.8 | 32.7 | 32.2 | 612 |
| Multidisciplinary | 24.3 | 54.4 | 36.1 | 31.4 | 41.7 | 500 |
| Biochemistry, Genetics and Molecular Biology | 23.4 | 17.0 | 69.4 | 61.1 | 64.3 | 1,814 |
| Medicine | 22.1 | 17.0 | 57.4 | 48.5 | 51.8 | 1,544 |
| Environmental Science | 20.2 | 17.2 | 37.3 | 45.6 | 33.9 | 800 |
| Earth and Planetary Sciences | 20.2 | 12.4 | 27.7 | 28.6 | 24.3 | 800 |
| Computer Science | 16.2 | 10.1 | 44.5 | 64.9 | 48.1 | 1,000 |
| Social Sciences | 15.6 | 5.1 | 17.8 | 41.3 | 8.8 | 830 |
| Mathematics | 14.4 | 9.4 | 37.0 | 45.0 | 31.5 | 980 |
| Physics and Astronomy | 14.0 | 17.3 | 48.2 | 46.0 | 41.3 | 1,225 |
| Materials Science | 13,9 | 9.4 | 4.,3 | 52.3 | 37.4 | 1,000 |
| Chemistry | 13.8 | 6.3 | 46.1 | 40.5 | 41.5 | 1,085 |
| Engineering | 13.4 | 5.9 | 41.1 | 46.1 | 33.4 | 600 |
| Arts & Humanities | 12.9 | 1.4 | 9.1 | 29.1 | 4.3 | 304 |
| Economics, Econometrics and Finance | 11.8 | 2.1 | 27.3 | 34.0 | 9.8 | 325 |
| Psychology | 11.5 | 2.6 | 27.5 | 38.7 | 16.7 | 841 |
| Business | 7.6 | 0.0 | 30.0 | 39.3 | 20.4 | 325 |

Table 3 indicates a large variation in OA adoption between disciplines. Outlets of better academic quality, as indicated by a top 10% SJR ranking, are for most disciplines less likely to be OA journals. However, in Physics and Astronomy, and particularly in the Multidisciplinary group, the share of OA is greater within the journals ranked in the top 10%. Highly ranked OA journals are quite rare in Psychology and Arts & Humanities, and next to non-existent in Economics and Finance, as well as Business. Commercial publishers dominate in the biomedical sciences, and biochemistry, where also born-OA journals and journals that charge APCs are common. Publishing charges are relatively rare in the social sciences and arts & humanities.

## Developments in different disciplines

In order to better understand why OA has spread at different speeds in different branches of science, it is necessary to look at the contexts of publishing in different disciplines and the historical developments of OA publishing. We concentrate on the fields that have received the most coverage of OA developments in the existing literature and which offer contrasting evolution paths.

### Medicine and Biochemistry

Professional publishers who specialize in OA journals first appeared in biomedicine. A number of factors may have influenced this development. Firstly, researchers in fields such as biomedicine are often capable of paying substantial publishing fees in the order of 2,000 to 3,000 USD, given the typical size of research budgets and external funding. Also, several major research funders like the National Institute of Health (USA) and Wellcome Trust (UK) early on defined quite strong policies requiring grantees to publish articles or manuscript versions OA. The NIH policy, which requires posting of open manuscript versions within 12 months of formal publication, influenced the decisions of many society journals in the biomedical field to opt for delayed OA of all articles after short embargo periods.

The two leading OA publishers, Public Library of Science (PLOS) and BioMedCentral (BMC), have differing origins. PLOS was the spin-off of a petition calling for scientists to stop submitting to journals that refused to make the full text articles available OA immediately or after a short delay. PLOS is a non-profit foundation and was started with a 9 million dollar grant from the Gordon and Betty Moore Foundation. The first journals were launched by PLOS in 2003. BMC was a purely commercial enterprise funded with venture capital. Other examples of successful OA publishers in biomedicine are Medknow from India and New Zealand-based Dove Medical Press. BMC, Medknow and Dove have all later been acquired by major commercial publishers.

A notable case of a converted journal is *Nucleic Acids Research*. When Oxford University Press made a strategic decision to start experimenting with both full and hybrid OA, they "flipped" this high impact and high volume journal to APC funded Open Access in 2005 [27].

Important to the relative success of newly founded OA journals in biomedicine is that some of the journals have had very substantial funding from the start and have strived for very high quality articles. *PLOS Biology* included Nobel Prize winners on the editorial board when they were launched. Howard Hughes Medical Institute, Max Planck Society and Wellcome Trust founded *eLife* in 2012 with seed funding of 25 million USD, and the journal aims at becoming an OA alternative to *Science* and *Nature*[28].

As a result of these developments 17% of the top decile of Scopus Journals in both Medicine and Biomedicine are Open Access, as indicated by Table 3.

**Multidisciplinary journals**

Although the category of multidisciplinary journals contains some journals covering all sciences, an important part of articles in these journals are in biomedicine. In addition to several established journals like *Nature*, *Science* and the *Proceedings of the National Academic of Sciences* (PNAS), the new category of mega-journals has emerged, thanks to the OA business model. Public Library of Science with *PLOS ONE* has been the forerunner in pioneering this new type of journal. Mega-journals carry out a novel type of peer review in which the scientific credibility of articles is vetted, but not the perceived significance of the results [29]. Such journals scale easily to even thousands of articles per year since paper space is not a restricting factor in electronic only publishing. Leading megajournals have established their credibility and they have achieved impact factors that are competitive for their fields. Megajournals attract authors not only with immediate OA, but also with rapid publication and predictable peer review[30]. In 2017, nineteen such OA megajournals published around 58,000 articles[31]. Due to this, the overall OA article share in the "multidisciplinary" category is 75%.

**Physics and Astronomy**

Within Physics, the subfield of High-Energy Physics pioneered Open Access in the 1990s, not by founding of OA journals, but rather by establishing a server for preprint manuscript versions[32]. The goal was to distribute new findings quickly and thus sustain evolution of research in the field. A tradition of exchanging such preprints existed previous to the Internet, and since many manuscripts are authored by large groups of scientists working in research organizations like CERN, the manuscripts have typically already undergone quite rigorous internal review. The arXiv server today hosts over 1.5 million manuscripts, and for researchers working in the field, it is the main forum for "publishing" their research results, while later formal journal publishing is mainly important for the record. In the wake of the success of arXiv, the leading research laboratories in the world created a consortium to collect pledges, diverting their subscription budgets for the leading journals to APC funding for the same journals. After several years of negotiations, the Sponsoring Consortium for Open Access Publishing in Particle Physics (SCOAP3) eventually reached a deal in 2014 with some leading publishers[33]. Four leading journals from Elsevier and Springer converted fully to OA, publishing 3,184 articles in 2017. In addition, all High Energy content in seven broader physics journals is also paid for in a hybrid arrangement. This factor explains both why the OA article share and the top SJR decile share is higher than the journal share, for physics and astronomy.

## Engineering and Computer Science

In many engineering disciplines and in computer science, publishing in conference proceedings is almost as important as publishing in journals [34], [35]. This assertion is supported by a typically high valuation of conference contributions in researcher evaluations. While conference proceedings account for 18% of all Scopus indexed serial publications, the balance is heavily tilted towards Engineering and Computer Science. Only 3% of publications in medicine are conference proceedings, but the same statistic is 40% for engineering disciplines and 65% for computer science. Unfortunately, we are not easily able to analyze the proportion of such conference series that are OA, since conference proceedings are not indexed by the DOAJ.

Looking at the data for OA journals only, Engineering and Computer Science are in the midrange of all disciplines, with 13.4 and 16.2 % journal OA shares, respectively. The shares of commercially published, journals (41% and 45%) are close to the average of 41% for all disciplines. Computer Science has the highest proportion of Born OA journals (64.9 %) of all disciplines, which may be explained by the object of study being rather young.

## Social Sciences, Arts and Humanities

The publishing patterns in the social sciences and in the arts and humanities differ significantly from those in the physical science disciplines. Publishing of book chapters or monographs is popular, and the "shelf life" of publications is often longer. This latter factor may be why a number of publishers have set longer embargo periods for green OA manuscript copies within these disciplines than for the physical sciences and biomedicine [36]. The relatively low shares of OA journals funded with APCs may reflect the more constrained availability of financing in these disciplines[37]. Scholarly societies and universities presses are the dominant publishers in Social Sciences, Arts and Humanities. Since such publishers often have just a single journal, many of them have used regional and national OA portals for the publishing. A good example is *Informaatiotutkimus (Information Research)*, published by the Society for Information Research on the Finnish Journal.fi portal. Other prominent examples of portals include Scielo and Redalyc in Latin America, Hrčak (Croatia), and Asian Journals On-line. For those journals that publish on their own software platforms, the open source OJS system has been a popular choice[38].

As evidenced in Figure 3 and Table 3, in the Social Sciences, Arts and Humanities both the share of OA journals published in the four leading countries as well as the share of OA journals among the top SJR decile are among the lowest.

In order to help both new and existing journals in the humanities, the recently launched Open Library of Humanities (OLH) uses a consortia model for funding its operations[39]. Due to over a hundred participating universities from around the world, publishing in the journals hosted by OLH is mostly free for authors.

**Business**

Out of all studied sub-disciplines, business has the lowest OA penetration (in journal and article shares, and also in the share of OA journals among top SJR-ranked journals). In addition to a high share of journals published in the Big four countries (see Figure 3), a potential factor influencing this outcome is the extremely entrenched position that established journals have in the journal rankings followed by business schools[40]. Rafols, et al. demonstrated how the UK Association of Business Schools journal ranking negatively influences business scholars wishing to publish in OA journals[41]. In their attempts to evaluate academic qualifications of business school faculties, the main international accreditations in business (EQUIS, AACSB, AMBA) put demands on research productivity, which is measured by publications in highly ranked selective journals, such as those included in the renowned Financial Times list of fifty top journals. This factor has likely contributed to the difficulty for newly founded OA journals to gain a foothold in these disciplines.

It is paradoxical that open access journals are much more common in information and library science (a subfield of social science) than in the thematically quite close Management Information Systems (a subfield of business), with shares of 20.1 % and only 5.1 %, respectively. The likely reason is that scholarly publishing and Open Access in itself is a topic for study and better known in the former field, whereas Management Information Systems is a subject taught at business schools and subject to the pressures for publishing in traditional journals with established rankings, as we discuss above.

## Discussion

Overall, this study confirms that the adoption of OA publishing is rather uneven across disciplines. This development is not surprising. Already Kling and McKing predicted this situation in their seminal article "Not Just a Matter of Time: Field Differences and the Shaping of Electronic Media in Supporting Scientific Communication"[42]. Subsequent studies have discussed how the research cultures within separate scholarly communities shape the uptake of open access[43],[44].

An important new contribution of this study is that it shows that the uneven pattern gets even more pronounced when we consider journals published in either the USA, UK, the Netherlands and Germany. The examination of journals published in the Big four countries (around 2/3) vs. in other countries is a simplified substitute for a much more detailed analysis of the types of publishers, their business strategies and current competitive situation and profitability. Those are important underlying factors explaining the uptake of OA (or, rather, the lack of uptake). The leading international publishers have so far had little incentive to change their predominant business model[45]. They have experimented with OA either on a small scale or they have entered the OA market by purchasing OA publishers that have already established themselves. Small society and university publishers, in turn, dominate the journals published in countries outside the Big four, and such journals (often published in languages other English) have had strong incentives to convert to OA, in particular if convenient and cheap solutions have been available in the form of collaborative OA portals. Geographic variation, nevertheless, cannot fully account for the variation across disciplines.

Commercial publishers have founded fewer OA journals in the social sciences and humanities, perhaps because they rely on APC income. Our results indicate very low percentages of APC funded OA journals for the social sciences (8.8%) and arts and humanities (4.3%). For instance Sage has tried to establish a megajournal for all the social sciences, *SAGE OPEN*, with only modest success.

An important factor, which is likely to have contributed to the uneven proliferation of OA across fields is if some OA journals have been able to reach top tier positions in their fields early on. This means that such journals are followed by leading researchers and promote OA journals as viable outlets for good quality articles as well as general awareness of OA and its benefits. Conversely newer OA journals have more trouble attracting the same attention or gaining similar traction.

Another interesting factor is the share of born OA versus converted journals in different fields. It is not surprising that the share of journals which have been OA from the start is highest in Computer Science (65 %), which is a young research field with new sub-specialties constantly emerging, closely followed by Biochemistry and related fields (61%). The lowest share of born OA is in the arts and humanities with 29 %.

## Conclusions and policy implications

Major research funders and ministries of education are influential stakeholders in the scholarly publishing ecosystem. Consequently, they have started to play a role in trying to influence the leading mainstream publishers to an accelerated transition to open access publishing. This is particularly true for Western and Northern European countries. Important milestones have been the UK Finch Report[46] and more recently Plan S [47], which has been gathering pledges from leading research funders from several countries to push for a more rapid adoption of OA.

A recent important development is also the push of several large library consortia to force major publishers to repackage their big subscription deals to include automatic APC payments for articles published by their faculty in hybrid journals. Such deals are called transformative agreements or "publish and read"[48]. There are already several in place, in particular as negotiated by the library consortia for countries like the UK, Netherlands, Germany, Austria and the Nordic countries. Also in North America University libraries have become active in this respect. The University of California has for instance cancelled its big deal license with Elsevier due to a breakdown in the negotiations[49]

In addition to being involved in big deal licensing negotiations university librarians across the world are also involved in many other OA related activities, where a knowledge of the current status of OA also across scholarly disciplines can be useful. In many universities the libraries handle OA advocacy. For instance Peter Suber, the Director of the Office for Scholarly Communication at the Harvard Widener Library, is an internationally leading OA expert. The libraries are often tasked with providing faculty with advice about both credible OA journals in their fields as well as predatory journals to avoid. In many universities the libraries manage the institutional repositories, which offer green alternatives to directly publishing in OA journals (MIT's open source repository D-SPACE has been widely adopted). Particularly in European Universities the libraries typically also manage the current research information systems, where faculty have to register all research publications. In our university, in order for the articles to count fully in research output assessments etc., in addition to registering the bibliographic meta-data, they have to either to be published in full OA journals, as paid OA articles in hybrid OA journals or the author has to upload a green OA manuscript copy in our institutional repository.

An important role for libraries can also be to act as intermediaries for paying APCs for OA articles (full or hybrid) to publishers. This often saves the authors all lot of administrative work and in some countries the universities can be refunded for the payments from central national funds. The UK provides a good example of such schemes[50].

In many instances university libraries also directly handle the publishing of OA journals. For instance *Information Research an International Electronic Journal*, which was founded by prof. Tom Wilson at the Sheffield University in 1995 is nowadays hosted and technically supported by the Lund University Libraries in Sweden. That same library has also otherwise played an active role in OA by starting the DOAJ index in 2003.

All in all, OA has had a slow start for almost two decades and has not yet turned into the mainstream business model for scholarly journal publishing. The subscription model has continued to be profitable and, thus, offered low incentives for the dominant publishers to switch their business models, and OA has so far mainly developed in fringe areas of the market and in selected research disciplines where the conditions for uptake have been favorable. Nevertheless, there are now strong on-going initiatives that are pushing for change, and OA might soon reach a critical mass, where developments could accelerate rapidly.

## Notes

[1] Peter Suber, "Open Access", MIT Press, Cambridge, Mass., 2012, available online at at https://mitpress.mit.edu/books/open-access [accessed 11 November 2019].

[2] Robert Merton, "The Normative Structure of Science", in Merton, Robert K. (ed.), The Sociology of Science: Theoretical and Empirical Investigations, University of Chicago Press, Chicago., 1973.

[3] Jonathan Tennant, François Waldner, Damien Jacques, Paola Masuzzo, Lauren Collister and Chris Hartgerink, "The academic, economic and societal impacts of Open Access: an evidence-based review", version 3, *F1000 Research* 5 (2016): 632, available online at https://f1000research.com/articles/5-632/v3 [accessed 11 November 2019].

[4] Erin McKiernan, Philip Bourne, Titus Brown et al, 2016. "Point of View: How open science helps researchers succeed". *eLife* 5 (2016): e16800, doi: 10.7554/eLife.16800

[5] Stevan Harnad, Tim Brody, Francois Vallieres, Les Carr, Steve Hitchcock, Yves Gingras, Charles Oppenheim, Heinmrich Stamerjohanns and Eberhard Hilf, "The access/impact problem and the green and gold roads to open access", *Serials Review* 30, No. 4 (2004): 310–314.

[6] Mikael Laakso and Bo-Christer Björk, "Delayed open access – an overlooked high-impact type of openly available scientific literature", *Journal of the American Society for Information Science and Technology* 64, No.7 (2013): 1323-1329.

[7] David Solomon, Mikael Laakso, Bo-Christer Björk and Peter Suber, "Converting Scholarly Journals to Open Access: A Review of Approaches and Experiences", Report, Harvard Library, Cambridge, Mass., 2016, available online at http://digitalcommons.unl.edu/scholcom/27/ [accessed 11 November 2019].

[8] Christine McVeigh, "Open Access Journals in the ISI Citation Databases: Analysis of Impact Factors and Citation Patterns - A citation study from Thomson Scientific", Report, ISI, 2004, available online at http://ips.clarivate.com//m/pdfs/openaccesscitations2.pdf [accessed 11 November 2019].

[9] David Solomon, Mikael Laakso and Bo-Christer Björk, "A longitudinal comparison of citation rates and growth among open access journals", *Journal of Informetrics* 7, No. 3 (2013): 642-650.

[10] Michael Jubb, Andrew Plume, Stephanie Oeben et al., "Monitoring the Transition to Open Access – December 2017", Report, Universities UK, 2017, available online at https://www.universitiesuk.ac.uk/policy-and-analysis/reports/Documents/2017/monitoring-transition-open-access-2017.pdf [accessed 11 November 2019].

[11] John Bohannon, "Who's afraid of peer review", *Science* 342, No. 6154 (2013): 60-65.

[12] Cenyu Shen and Bo-Christer Björk, "'Predatory' open access: a longitudinal study of article volumes and market characteristics", *BMC Medicine* 13, (2015): e230, available online at doi: 10.1186/s12916-015-0469-2 [accessed 11 November 2019].

[13] Abel Packer et al., "SciELO - 15 Years of Open Access: an analytic study of Open Access and scholarly communication", UNESCO, Paris, 2014, available online at doi: http://dx.doi.org/10.7476/9789230012373 [accessed 11 November 2019].

[14] Mikael Laakso and Bo-Christer Björk, "Anatomy of open access publishing: a study of longitudinal development and internal structure*", BMC Medicine* 10 (2012), e214, available online at https://doi.org/10.1186/1741-7015-10-124 [accessed 11 November 2019].

[15] Mohammadamin Erfanmanesh, "Status and quality of open access journals in Scopus", *Collection Building* 36, No. 4 (2017): 155-162.

[16] Yassine Gargouri, Vincent Larivière, Yves Gingras, Les Carr and Stevan Harnad, "Green and gold open access percentages and growth, by discipline" ArXiv preprint, 2012, available online at https://arxiv.org/abs/1206.3664 [accessed 11 November 2019].

[17] Éric Archambault, Didier Amyot, Philippe Deschamps, et al.,"Proportion of Open Access Papers Published in Peer-Reviewed Journals at the European and World Levels—1996–2013", Report for the European Commission, Science Metrics, Montreal, Canada, 2014, available online at http://digitalcommons.unl.edu/scholcom/8 [accessed 11 November 2019].

[18] Heather Piwowar, Jason Priem, Vincent Larivière et al., "The state of OA: a large-scale analysis of the prevalence and impact of Open Access articles", PeerJ, 6 (2018): e4375, available online et doi 10.7717/peerj.4375 [accessed 11 November 2019].

[19] Laakso and Björk, "Anatomy of open access publishing: a study of longitudinal development and internal structure", *BMC Medicine* 10 (2012), e214, available online at https://doi.org/10.1186/1741-7015-10-124 [accessed 11 November 2019].

[20] Solomon, Laakso and Björk, "A longitudinal comparison of citation rates and growth among open access journals", *Journal of Informetrics* 7, No. 3 (2013): 642-650.

[21] Erfanmanesh, "Status and quality of open access journals in Scopus". *Collection Building* 36, No. 4 (2017): 155-162.

[22] Nathalie Cornic, "ROAD: the Directory of Open Access Scholarly Resources to Promote Open Access Worldwide, Elpub 2016 conference presentation, available online at https://elpub.architexturez.net/system/files/120_elpub2016.pdf [accessed 11 November 2019].

[23] Rob Johnson, Anthony Watkinson and Michael Mabe, "The STM Report, An overview of scientific and scholarly publishing", fifth edition, October 2018, The International Association of Scientific, Technical and Medical Publishers, available online at https://www.stm-assoc.org/2018_10_04_STM_Report_2018.pdf [accessed 11 November 2019].

[24] Ladislava Suchá and Jela Steinerevá, "Journal publishing models in the Czech Republic", *Learned Publishing* 28, No. 4 (2015): 239-249.

[25] Miguel Navas, "Open Access Journals in Spain", DOAJ blogpost, 8.12.2016, available online at https://blog.doaj.org/2016/12/08/open-access-journals-in-spain/ [accessed 11 November 2019].

[26] Solomon, Laakso and Björk, "A longitudinal comparison of citation rates and growth among open access journals", *Journal of Informetrics* 7, No. 3 (2013): 642-650.

[27] Claire Bird, "Continued adventures in open access: 2009 perspective", *Learned Publishing* 23, No. 2 (2010): 107-116.

[28] Ewen Callaway, "Open-access journal eLife gets £25-million boost". *Nature* 534, No. 7605 (2016): 14–15, available online at doi:10.1038/534014a [accessed 11 November 2019].

[29] Simon Wakeling, Peter Willett, Claire Creaser, Jenny Fry, Stephen Pinfield and Valérie Spezi, "Open-Access Mega-Journals: A Bibliometric Profile", *PLOS ONE* 11 (2016): e0165359, available online at doi: 10.1371/journal.pone.0165359 [accessed 11 November 2019].

[30] David Solomon, "A Survey of authors publishing in four megajournals", *PeerJ* 2 (2014): e365, available online at doi: 10.7717/peerj.365 [accessed 11 November 2019].
 [accessed 11 November 2019].

[31] Bo-Christer Björk, "Evolution of the scholarly mega-journal, 2006–2017", PeerJ 6 (2018):e4357, available online at doi: 10.7717/peerj.4357 [accessed 11 November 2019].

[32] Tracy Vence, "One Million Preprints and Counting", *The Scientist*, 29.12.2014, available online at https://www.the-scientist.com/daily-news/qa-1-million-preprints-and-counting-36168 [accessed 11 November 2019].

[33] Richard Van Noorden, "Particle-physics papers set free", *Nature* 505, No. 141 (2014), Available online at doi:10.1038/505141a [accessed 11 November 2019].

[34] Carol Tenopir and Donald King, "Communication Patterns of Engineers", Wiley-IEEE Press, Piscataway, NJ, USA, 2004.

[35] Jinseok Kim, "Author-based analysis of conference versus journal publication in computer science", *Journal of the Society for Information Science and Technology* 70, No. 1 (2019): 71-82.

[36] Mikael Laakso, "Green open access policies of scholarly journal publishers: a study of what, when, and where self-archiving is allowed", *Scientometrics*, 99, No. 2 (2014): 475-494.

[37] David Solomon and Bo-Christer Björk, "Publication Fees in Open Access Publishing: Sources of Funding and Factors Influencing Choice of Journal", *Journal of the American Society for Information Science and Technology*, vol. 63, no. 1 (2012): 98-107.

[38] Brian Edgar and John Willinsky, "A survey of scholarly journals using open journal system", *Scholarly and Research Communication* 1, No.2 (2010), available online at http://src-online.ca/index.php/src/article/view/24/ [accessed 11 November 2019].

[39] Martin Eve, "All That Glisters: Investigating Collective Funding Mechanisms for Gold Open Access in Humanities Disciplines", *Journal of Librarianship and Scholarly Communication* 2, No. 3 (2014), eP1131, available online at https://jlsc-pub.org/articles/abstract/10.7710/2162-3309.1131/ [accessed 11 November 2019].

[40] Nancy Adler and Anne-Will Harzing, "When Knowledge Wins: Transcending the Sense and Nonsense of Academic Rankings", *Academy of Management Learning & Education* 8, No. 1 (2009): 72–95.

[41] Ismael Rafols, Loet Leydesdorff, Alice O'Hare, Paul Nightingale and Andy Stirling, "How journal rankings can suppress interdisciplinary research: A comparison between Innovation Studies and Business & Management", *Research Policy* 41, No. 7 (2012): 1262-1282.

[42] Rob Kling, Geoffrey McKim, "Not Just a Matter of Time: Field Differences and the Shaping of Electronic Media in Supporting Scientific Communication", *Journal of the American Society for Information Science* 51, No.14 (2000): 1306-1320.

[43] Jenny Fry and Sanna Talja, "The cultural shaping of scholarly communication: Explaining e-journal use within and across academic fields", In ASIST 2004: Proceedings of the 67th ASIST

Annual Meeting, edited by Linda Schamber and Carol Barry, Information Today, Medford , NJ: 20-30.

[44] Anna Severin, Matthias Egger, Martin Eve and Daniel Hürlimann, "Discipline-specific open access publishing practices and barriers to change: an evidence-based review", [version 1;], *F1000Research* 7, (2018): 1925. Available online at https://doi.org/10.12688/f1000research.17328.1 [accessed 11 November 2019].

[45] Bo-Christer Björk, "Scholarly journal publishing in transition– from restricted to open access", *Electronic Markets, The International Journal on Networked Business* 27, No. 2 (2017): 101-109.

[46] Janet Finch, "Accessibility, sustainability, excellence: how to expand access to research publications", Report of the Working Group on Expanding Access to Published Research Findings, The Association of Commonwealth Universitities, UK, 2012, available online at http://www.researchinfonet.org/publish/finch/ [accessed 11 November 2019].

[47] Holly Else, "Radical open-access plan could spell end to journal subscriptions", *Nature* 561. No. 7721 (2018): 17–18, available online at DOI: 10.1038/d41586-018-06178-7 [accessed 11 November 2019].

[48] JISC, "Requirements for transformative Open Access agreements: Accelerating the transition to immediate and worldwide Open Access", Report, JISC collections, UK, 2019, available online at https://www.jisc-collections.ac.uk/Transformative-OA-Reqs/ DOI: 10.1038/d41586-018-06178-7 [accessed 11 November 2019].

[49] Ivy Anderson et al. "Fact Check: What you may have heard about the dispute between UC and Elsevier", Office of Scholarly Communication, University of California, August 2nd 2019, available online at https://osc.universityofcalifornia.edu/2019/08/fact-check-uc-and-elsevier/ [accessed 11 November 2019].

[50] Stephen Pinfield, Jeniffer Salter and Peter Bath; "The Total Cost of Publication in a Hybrid Open-Access Environment: Institutional Approaches to Funding Journal Article-Processing Charges in Combination With Subscriptions", *Journal of the Association for Information Science and Technology* 67, No. 7 (2016):1751–1766.